# DPN – Dependability Priority Numbers

Zhensheng Guo, Marc Zeller

Siemens AG, Corporate Technology
Otto-Hahn-Ring 6, 81739 Munich, Germany
`joe.guo@siemens.com`
`marc.zeller@siemens.com`

**Abstract.** This paper proposes a novel model-based approach to combine the quantitative dependability (safety, reliability, availability, maintainability and IT security) analysis and trade-off analysis. The proposed approach is called DPN (Dependability Priority Numbers) and allows the comparison of different actual dependability characteristics of a systems with its target values and evaluates them regarding trade-off analysis criteria. Therefore, the target values of system dependability characteristics are taken as requirements, while the actual value of a specific system design are provided by quantitative and qualitative dependability analysis (FHA, FMEA, FMEDA, of CFT-based FTA). The DPN approach evaluates the fulfillment of individual target requirements and perform trade-offs between analysis objectives. We present the workflow and meta-model of the DPN approach, and illustrate our approach using a case study on a brake warning contact system. Hence, we demonstrate how the model-based DPNs improve system dependability by selecting the project crucial dependable design alternatives or measures.

## 1      Introduction

Ref. [9] defines *dependability of a system is the ability to avoid service failures that are more frequent and more severe than is acceptable* and it contains the following properties: *safety*, *reliability*, *availability*, *integrity* (security), and *maintainability*. Dependability trade-off analysis is basically the analysis of dependencies and conflicts between dependability properties according to the fulfillment of targets and to make trade-offs among these properties [1][2][8][11][13]. Quantitative dependability analysis deals with quantitative analysis of safety, reliability, availability, maintainability and security properties of a system design. Examples are Failure mode Effect Diagnostic Analysis (FMEDA), Fault Tree Analysis (FTA) etc. Currently the trade-off analysis of the dependability properties assumes in many cases that the target values to be fulfilled by the design alternatives, and actual values that the design alternatives hold, are given. Based on these values, acceptable limits and evaluation criteria, trade-off analyses are performed. However, the actual quantitative values of dependability properties of design alternatives in many cases are not given and need to be obtained. The techniques to perform (model-based) quantitative dependability analysis and to perform trade-off analysis are usually performed separately, or in other words, they are not combined sufficiently for effective quantitative dependability trade-off analysis.

In this work we describe with Dependability Priority Numbers (DPN) an approach to combine these two engineering fields and show how model-based quantitative dependability analysis techniques such as Component Fault Trees [10] can help to perform dependability trade-off analysis.

This paper is arranged in the following sections: Section 2 provides an overview of related work, Section 3 illustrates an approach, which is named Dependability Priority Number (DPN); Section 4 shows a case study on a brake warning contact system; Section 5 concludes this paper.

## 2  Related Work

Typically, the comparison of different design alternatives is the objective of dependability trade-off analysis. The design alternative that fulfills more dependability properties will be normally chosen as the solution. Today, there are some approaches to model the obtained dependability properties, e.g. through GSN [2], Modelica [6] etc., but the source of the quantitative value of the overall dependability is seldomly handled.

Ref. [1] uses vulnerability attack graph and goal graph to determine the dependencies between the security goals and tasks. This method mentions the use of trade-off analysis parameters such as risk acceptance criteria, standards, laws, regulations, policies, stakeholder goals, budget, and time-to-market. Ref. [2] utilizes DDA (Dependability Deviation Analysis) and GSN (Goal Structuring Notation) to perform trade-off analysis. This method uses GSN with acceptable limits to model the fulfillment of the design alternatives under certain scenarios. Ref. [3] emphasizes the role of scenarios and upper and lower bounds of acceptable limits in the trade-off analysis that is illustrated in [2]. Ref. Ref. [4] proposes a quantitative estimation method of the different dependability properties, in which expert estimations of the fulfillment of dependability properties are used. Ref. Ref. [5] uses an UML extension to describes the dependability properties and uses Deterministic and Stochastic Petri Net to perform dependability modelling. Ref. [6] uses Modelica and Bayesian Network simulation to identify the violence of the dependability requirements. Ref. [7] presents a trade-off analysis procedure to prioritize the different dependability requirements.

Ref. [7][11][12] proposed formulas to calculate the utility or value function of dependability of individual design alternatives. Ref. [7] uses product of weight and values function results to calculate the evaluation result of dependability properties such as performance, security and fault tolerance. For the calculation they use the following formulas:

$$evaluation\ result = max \frac{1}{n}[\sum_{i=1}^{n} v_i]\ with\ v_i \geq v_{min}\ \ for\ without\ weight$$
$$max \frac{1}{n}[\sum_{i=1}^{n} a_i v_i]\ with\ v_i \geq v_{min}\ and\ \sum a_i = 1, a_i > 0\ \ for\ with\ weight \quad (1)$$

Ref. [11] defines the dependability properties evaluation results as $x_i$ and takes the sum of value function of $x_i$ as the result of the overall dependability value. In addition, they use the sum of the products of the weights of the individual properties and their evaluation results of $x_i$ as the dependability value. The authors argue that the sum of the weights of dependability properties shall be 1:

$$v(x_1, x_{2,\ldots,}x_n) = v(x_1) + v(x_2) + \cdots + v(x_n) = \sum_{i=1}^{n} v(x_i)$$
$$\text{or}$$
$$v = \sum_{k=1}^{n} w_i x_i \ \ with \ w_i \geq 0 \ and \ \sum w_i = 1 \tag{2}$$

The decision-making procedure according to this work includes the following steps: identification of the subjective such as design alternatives; definition of the analysis criteria; Performance of the evaluation; selection of the value function and determination of combinable criteria. The precondition of the combining the criteria is that the criteria are mutual independent, and it is possible to determine the final equation for calculating the value of fulfillment of dependability properties. Ref. [1][12] proposed the following essential definitions for dependability evaluation: Preference function based on certainty (such as probability) is defined as value function, preference function based on risk (such as weights) is defined as a utility function. In [12], weights of a criteria/properties $w_{(i)}$ and value of this criteria $v_{(i)}$ are used to calculate utility of alternatives:

$$v = \sum_i w_{(i)} v_{(i)} \quad \text{or} \quad v = \sum_i p_{(i)} v_{(i)} \ \ where \ p \ denotes \ probability \tag{3}$$

Ref. [8] illustrates an approach by use of GSN and its evaluation process to perform the trade-off analysis of dependability properties. The following aspects are essential to perform the trade-off analysis for this survey: goals of stakeholders; function for scenarios; related dependability properties; target value of dependability properties; traceability to the requirements; acceptance criteria; determination of compromise region. According to their work, the scenarios (consist of stimuli, responses) and target / limit are essential for performing trade-off analysis. However, in this paper, the use of the dependability analyses is not illustrated in detail. Ref. [13] handles the trade-off analysis in a very thorough way. They proposed the following processes: identification of the concern of trade-analysis; definition of the deviation and failures; derivation of dependability requirements; identification of goals, target and limits; identification of alternatives; identification of trade-off argument based on GSN; evaluation of alternatives and decision making. The evaluation of the alternatives is done based on evaluation of the related criteria. The final value is produced with consideration of the weight. Matrix calculation is used for this evaluation process. The qualitative safety analysis techniques such as Hazard and Operability Analysis (HAZOP), Failure Mode and Effect Analysis (FMEA) are used for identifying the failures and further the dependability requirements. However, such analysis techniques are not reused to analyze the alternatives and the model-based quantitative safety analysis is not used in their work. [14] proposes a method to address the cost-benefit trade-off analysis. The following evaluation criteria are considered as essential: priorities; standards; laws; regulations; business goals; budget; policies. Taking a retrospective look at the related works, we can draw the conclusion that the dependability trade-off analysis were performed without integrating the model-based quantitative dependability analysis techniques.

## 3  Dependability Priority Numbers

In this section, we present the concept of *Dependability Priority Numbers (DPN)*. First, the result of this approach and its formula are described. Afterwards the workflow of DPN analyses is presented in detail. Moreover, the metamodel and its usage will be depicted.

By introducing a Dependability Priority Number, analysis object is extended from design alternatives to at least alternatives and the measures for mitigating hazard or risk will be analyzed. They will be analyzed qualitatively and/or quantitatively towards an overall result of the quality of the system in terms of dependability. The overall fulfillment of the dependability properties is presented by comparing the actual and expected DPN and also by comparison between the actual DPNs. The conflicts and dependencies between the dependability properties will be identified or solved during this process implicitly.

In this work, we use first the concept of weights to calculate the overall dependability value. Therefore, the utility values will be calculated according to the definition in [11]. However, the calculation of DPN can also be based on risk / probability. The result of the calculation of the utilities/values of the alternatives is named the *Dependability Priority Number (DPN)* (instead of using the rather general term, Utility or Value.). Because the result deals in deed with the prioritization of the alternatives, and this prioritization has certain similarity with the *Risk Priority Number*. Based on [11][7][12], the following formula is derived:

$$DPN_j = \sum_{i=1}^{n} X_{ij} * K_i \qquad (4)$$

Where
$n$ : number of the dependability properties;
$X_{ij}$: Evaluation result, correlates with acceptance level. If $X_{ij}$: 0: totally unacceptable, 1: totally acceptable. "i" for the index of dependability properties, "j" for alternatives / measures;
$K_i$: weight (or probability) coefficient of the individual dependability properties, according to the importance of current dependability properties. $\Sigma_i K_i$ not necessarily equals to 1

DPN uses a slightly changed formula of (1), (2) and (3) which are presented in Section 2. The $w_i$ or $a_i$ is replaced by weight (or probability) coefficient $K_i$, basically they are all the weights (except that $k_i$ can contain probability additionally). The difference of $K_i$ and $w_i$ or $a_i$ is that the sum of the weight coefficients $K_i$ used for DPN is not necessarily 1, this definition has the benefit for tracing back the causing property intuitionally in case of changing of overall DPNs. This means, that if the DPN is changed for example from 109.11 to 111.11, (assume the utilized weights are 100, 10, 1, 0.1 and 0.01 for safety, reliability, availability etc.) we know therefore in this case there is an improvement on the reliability (improvement on the second digit). The weight $K_i$ are generally determined by the domain expert according to the importance of the dependability properties. The selection of weights follows additionally the rule of distinguishing dependability properties big enough so that the weights of properties do not counterweight in case value changes. The weights can also be derived based on results of dependability analysis such as RPN out of FMEA or failure rates out of FTA. The result of DPN as simple numbers offers an intuitive and direct way to represent the overall fulfillment of the dependability goal and to compare variants.

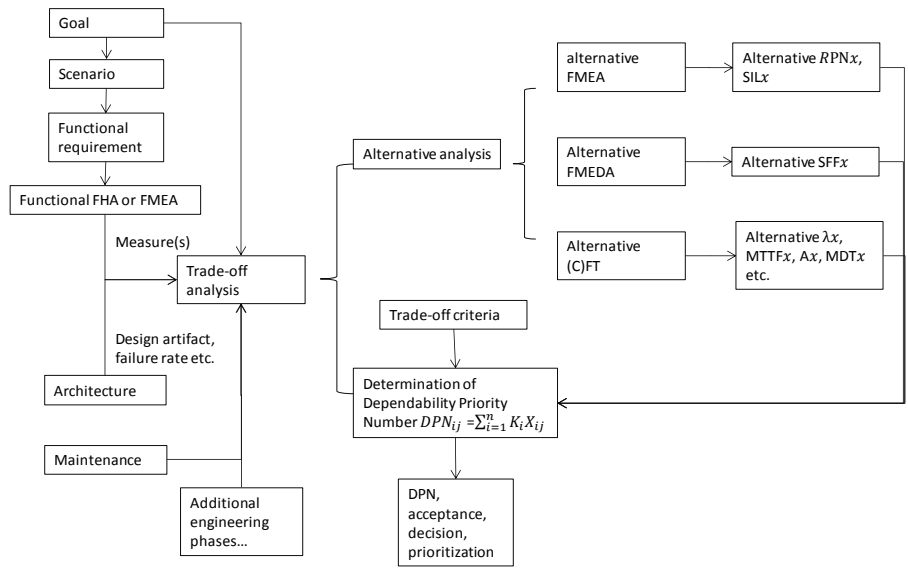

**Fig. 1.** Workflow to determine Dependability Priority Number

In Fig. 1. the workflow for determining Dependability Priority Numbers is illustrated. This workflow contains:

1. Elicitation of the goals of the stakeholders. Here the typical goal graph methods, such as GSN [2], i* [1] for Non-Functional Requirements etc. can be used. A coarse trade-off analysis among the identified goals can be performed, in order to identify the possible limits, dependencies and conflicts.
2. Based on the identified goals, the relevant scenarios with certain execution sequences will be determined. An example of such scenarios is robot x shall be stopped when safety bumper is engaged. Scenarios define the aims and scope of the trade-off analysis.

3. Typically, the functional requirements will be elicited based on the identified scenarios. If there are no standardized requirements and their THR, the functional requirements are to be elicited for the specific project.
4. Based on the identified functional requirements, the Functional Hazard Analysis (FHA) or function-based Failure Mode and Effect Criticality Analysis (FMECA) will be performed. The corresponding hazards, their Risk Priority Numbers (RPN), their Safety Integrity Level (SIL), and available measures will be identified. For fulfilling the predefined multiple quality goals (e.g. SIL) additional measures are to be identified. Traditionally only one measure is identified for fulfilling the predefined quality goal. By using DPN multiple measures will be identified by use of the dependability analysis repeatedly.
5. Trade-off analysis will be performed among alternative measures. If there are no further information about the system components and their failure rates, the qualitative FMEA or Functional Hazard Analysis (FHA) will be performed repeatedly, where the improvements of the quality in SIL or RPN of the alternative could be compared with the original (first) measure. The possible conflicts to other dependability properties could be identified by observing the interchanging of DPNs. In these steps of trade-off analysis, the expert estimation is required. The following trade-off analysis is to be performed based on the trade-off criteria (based on [2][8][11][13][14]):

    o Determination of actual value of dependability properties $v_a$;
    o Determination and comparison of target/expected value $v_e$ with $v_a$;
    o Determination and comparing of acceptable upper / lower limit with $v_a$ ;
    o Evaluation of the benefit of actual better value e.g. ($v_a \geq v_e$) / drawback of actual worse value e.g. ($v_a < v_e$);
    o Determination of the cost of improvement towards expected value e.g. ($v_a < v_e$);
    o Determination of time-to-achievement of the improvement e.g. ($v_a < v_e$);;
    o Determination of overall acceptance $X_{ij}$;
    o Derivation of further action

6. The actual value in the trade-off criteria could be obtained by FHA, Risk Priority Number through FMECA qualitatively or quantitatively by the FMEDA, (Component) Fault Tree Analysis (FTA) or Fault Tree Analysis (FT) or other quantitative dependability techniques.

The results of such dependability assessments / analyses will be used for the rest of quantitative dependability trade-off analysis: Failure rate $\lambda$ and SIL for the safety property, Mean Time Between/To Failure (MTBF/MTTF) for the reliability property, Availability value for the availability property, Mean Down Time for the maintainability etc. After determining measures and alternatives, they are modelled by a model-based (Component) Fault Tree. The results of these analyses are then compared between each of the system design alternatives. For Safety the calculated failure rate $\lambda$ and even qualitative RPN, SIL are used as „actual value", "expected value" is typically predefined either by the authorities or by the references systems.

By using FMEDA for determining Safe Failure Fraction (for estimation of the Safety Integrity Level) and dangerous undetected failures, the FMEDA will be performed several times according to the number of alternatives. The calculated SFFs, failure rates and the corresponding SILs will be then be used as actual value for the trade-off analysis. In case the new measure neither leads to architecture changes, nor to a structural update in the fault tree, the changed availability can still be captured by e.g. the changed Mean Down Time. For example, if stopping the train in case of warning contact "high" (warning contact is responsible for worn out status of the brake), affects the availability too negatively (unacceptable) and the measure of "stop" has no remarkable improvement of safety, in addition "low speed drive" is sufficient (regarding safety) to handle this warning contact. The "low speed" can then be used to replace "stop" as measure in case of warning contact "high". This change will obviously improve the availability of the train, and without compromise of the safety. This change does not necessarily change the fault tree structure of the train. But down time will be then reduced. The reduced down time will affect the calculation of availability positively because of $A = \frac{MTBF}{MTBF+MDT}$ for repairable systems. Through this way the availability comparison between the original solution "stop" and new solution "low speed drive" can be done even without changing the structure of the fault tree.

In DPN the quantitative analysis techniques such as the FTA and FMEDA are reused to calculate the influence of different alternatives on the overall system. Different system failure rates could be observed, because of different architectures or even different value of the parameter. The comparison of alternatives is performed regarding trade-off criteria.

Partially according to the industrial practice, there are for instance the following categories for the subjective trade-off criteria to be used for evaluating the alternatives:

- Benefit of the actual better value: None; Better life time cause of better quality; Better reliability or availability of the system; Potential reputation benefit; Eventually better sale price.
- Drawback of the actual worse value: None; No certificate; Financial disaster; Worse availability; Damage of reputation; Postpone of the project finish time; Increased purchase cost.
- Cost for improvement towards expected value: None; Ignorable; Proportional; Quite high; Too high.
- Time for achieving the expected value: None; Ignorable; Proportional; Quite long; Too long.
- Further action: None; Redundancy; Use of higher quality component; Development of new component.
- Acceptance level: 0: totally unacceptable; 0.2: almost unacceptable; 0.4: predominantly unacceptable; 0.6: predominantly acceptable; 0.8: almost acceptable; 1: totally acceptable

The overall acceptance (between 0 and 1) is represented by the value of $X_{ij}$, together with estimated value $K_i$. Based on these values DPNs are be calculated (according to (4)). Afterwards, the DPNs of different design alternatives are compared. The higher value means basically the better dependability. And the detailed comparison according to the single dependability properties can also be done. The comparison shall not only be done based on the subjective evaluation value, but also on the objective calculated value. Based on the such comparisons, the acceptance of the alternative can be determined. The mutual dependency, the conflicts are represented through the interchanging of evaluation (or calculated) values. For example if DPN changes from 111.10 to 110.11 directly, we know that there is a conflict between availability and security. Because increase of the security (from 0 to 1) causes decrease of the availability (from 1 to 0). DPN are calculated for instance in the following way: assume safety has the weight of 100, reliability has the weight of 10 and so on. And the $X_{ij}$ all have value "1" for totally acceptance. The expected Dependability Priority Number would be $DPN_{expected} = 100 * 1 + 10 * 1 + \cdots = 111.11$. This expected value is then used to compare with the actual values.

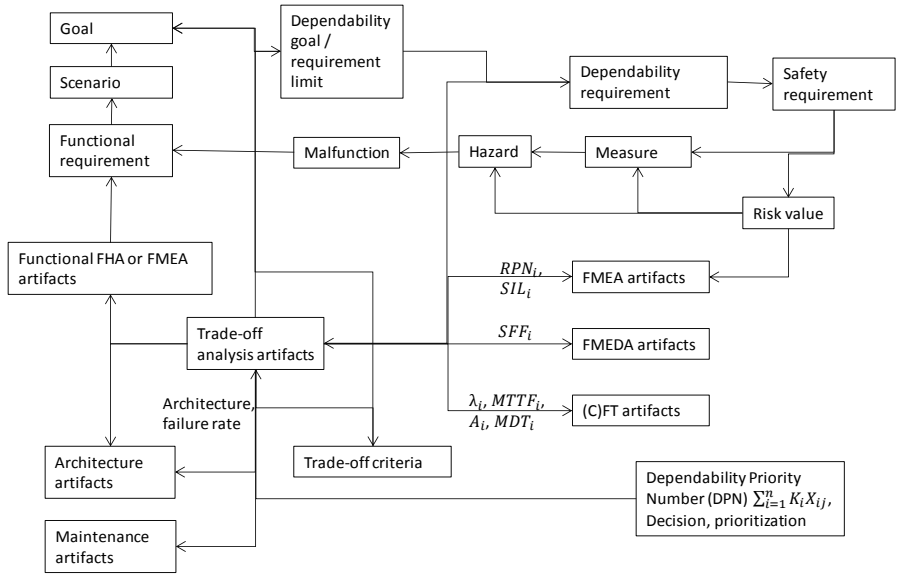

**Fig. 2.** Metamodel of the Dependability Priority Number

As illustrated in the Fig. 2. , goal, scenario, and functional requirements are the bases of the trade-off analysis and define the subjects of the trade-off analysis. During the model-based dependability analysis, the following data are identified step by step by use of this meta model:

- Malfunction, hazards are identified by use of e.g. FHA based on the functional requirement. The limit of goals can be used as limit of underlying requirements for the further trade-off analysis;
- Based on the hazard incl. its risk value the multiple measures are identified;
- The trade-off analysis of alternative measures can be qualitative or quantitative. Qualitative trade-off analysis can be the repeated model-based FHA or FMEA analyses for determining the reduced RPN or SIL by use of different measures. Such results are represented as $RPN_i$ and $SIL_i$. Where the i indicates the sequential number representing each of the design variants. Quantitative trade-off analyses are performed through repeated FMEDA or (C)FT for calculating the $\lambda_i$, $MTTF_i$/ $MTBF_i$, $A_i$, $MDT_i$. Through the comparison of the $\lambda_i$ and $\lambda_{i+1}$ the variant which is better in terms of safety or reliability can be identified. Further the comparison of aforementioned other values could contribute to an overall evaluation value of the dependability properties.

- The calculation of the expected and actual values are performed by the equation (4) based on the evaluation of the trade-off criteria as mentioned in the workflow section. The $DPN_{extected}$ and $DPN_{actual}$ are then used further to determine whether the $DPN_{actual} \geq DPN_{extected}$. If this is the case, all the dependability properties are fulfilled, otherwise a or some or even all the dependability properties are possibility not fulfilled. The not fulfilled dependability properties need basically further measure until this is fulfilled. In the end all the dependability properties shall be in general fulfilled. However, there can be conflicts by fulfilling the different properties, for example the fulfillment of safety properties means in certain circumstances the harm to the availability. This happens for example if a train is stopped for certain safety reason, but this means immediately the reduction of the availability. Compromise has to be made in this case. DPN result is shown at the bottom-right corner of Fig. 1. DPN approach consists of both the process of Fig. 1. and the data set of Fig. 2.
- Not only the measures, the quality goals and the functional requirements are the possible objects of the trade-off analysis but also the design artifacts and maintenance artifacts are also potential objects. Design artifacts offer among others the design alternative. Maintenance artifacts can be for instance the size of the maintenance team, possible maintenance strategy as conditions which also play roles in determining the maintenance priority number (basically the calculable Mean Down Time). By changes of dependa. goals, the DPN process shall be repeated totally or partially, according to the result of the similarity analysis between the old and new goals.

## 4    Case study – Brake Warning Contact

This section presents a case study form the railway domain based on a brake warning contact. The brake warning contact monitors the status of the brakes, if the thickness of brakes is detected less than allowed, a warning message will be sent to the dashboard, the train will be set to degraded mode. By using this example, the workflow of DPN is explained in detail:

1. Performing FMECA:
   The following functional requirement has been identified: If the warning contact is high, the warning contact sensor shall send the warning signal to the dashboard and set the train to degraded mode. Based on the identified function a FMECA is performed and multiple measures (redundancy and monitoring are identified in Table 1).

**Table 1.** FMECA inclusive multiple measures

| Measure | New RPN | New Probab. | New Detect. | New Sever. | Further Action |
|---|---|---|---|---|---|
| measure 1: Redundant warning contact sensor | 56 | 1 | 7 | 8 | no |
| Measure 2: Monitoring of warning contact | 16 | 1 | 2 | 8 | |

2. Performing FMEDA:

The FMEDA identifies the dangerous undetected failure rates of redundancy (5 fit, see Table 2). Moreover, the dangerous undetected failure rate of monitoring is 1, under the assumption that the monitoring detects 90% dangerous failure.

**Table 2.** Results of FMEDA for multiple measures

| Detection and control measure | Detection coverage (DC) | Failure rate of dangerous undetected | Failure rate of dangerous detected |
|---|---|---|---|
| Redundancy | 50% | 5 | 5 |
| Monitoring | 90% | 1 | 9 |

3. Performing Component Fault Tree Analyses for the following design alternatives (measures):

- **Without measure:** This fault tree contains only the components "power supply" and "brake warning contact" combined using an OR-gate.
- **With measure 1 of redundancy**: The component "brake warning contact" is doubled and because of the redundancy, the two instances are combined using an AND-gate. This subtree with the AND-gate is then combined with the "power supply" component using an OR-gate.
- **With measure 2 of monitoring (3 variants** with failure rate (FR) of 10000 fit, 10 fit, and 1 fit): As illustrated in Fig. 3, the use of the monitoring mechanism introduces additional failure possibility, because the monitoring can also fail. In this case, the brake warning contact fails if 1) the monitoring fails and the brake warning contact (9 fit) dangerous detectable fails or 2) brake warning contact dangerous undetected fails (1 fit). The failure rate of monitoring mechanism plays here a significant role. 10,000 fit, 10 fit and 1 fit are selected to perform this comparison in this case study. 8760 hours (1 year) was used as mission time, 24 hours were used as Mean Down Time of the basic events. Based on such data, CFT-based dependability / Reliability Availability Maintenance Safety (RAMS) properties are modelled and calculated. The modelling of the CFT is performed using ComposR, a Siemens-internal model-based safety and reliability analysis tool. The calculation is done using ZUSIM, a Siemens-internal safety and reliability calculation engines that has been used since decades.

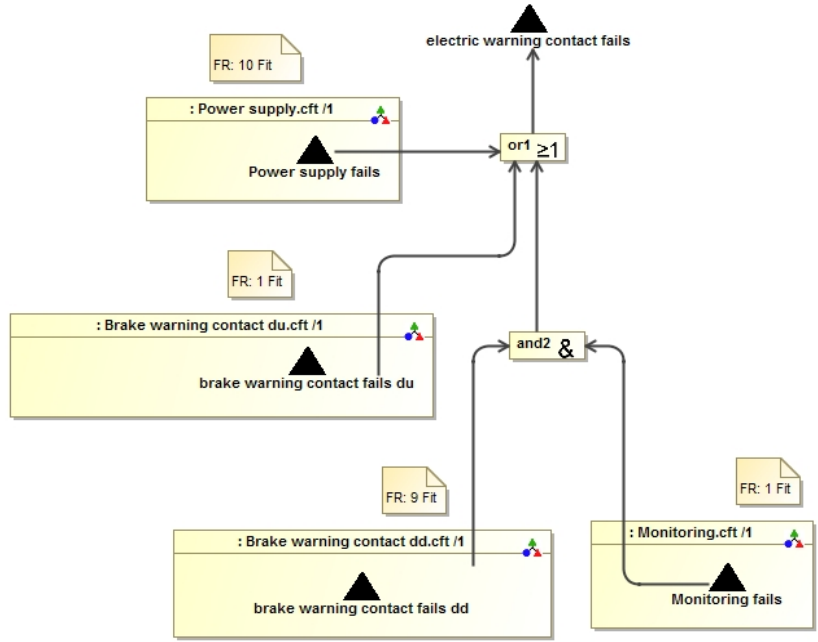

**Fig. 3.** Component Fault Tree of the measure monitoring

The goal of the quantitative analysis is to determine the measure which fulfills all (or more) the target values. In the CFT as depicted in Fig. 3, four components (power supply, brake warning contact dangerous undetected, brake warning contact dangerous detected, monitoring) and two gates (one AND- and one OR-gates) are modeled. The analysis results of all 5 design alternatives are summarized in Table 3. The individual analysis results (such as failure rate) are used as actual failure rate which serve as basis to be compared with the target/expected value. Other (reliability, availability etc.) actual and target values are also compared in the same way. The following formulas are used to calculate MDT by use of ZUSIM: for OR gate $MDT_{OR} = \frac{MTBF_1 * MDT_2 + MTBF_2 * MDT1_1}{MTBF_1 + MTBF_2}$ and for AND gate $MDT_{OR} = \frac{MDT_1 * MDT1_2}{MDT_1 + MDT_2}$.

**Table 3.** Summarized dependability calculation results of the measures by use of ZUSIM

| Result | Without measure | With redundancy | With monitoring FR: 10000 fit | With monitoring FR: 10 fit | With monitoring FR: 1 fit |
|---|---|---|---|---|---|
| Availability | 99,9999800000 00% | 99,999999 99999% | 99,999995% | 99,99999% | 99,999995% |

| Unavaila-bility | 2,40E-07 | 1,15E-13 | 4,81E-06 | 4,80E-06 | 4,80E-06 |
|---|---|---|---|---|---|
| MTBF (h) | 1,00E+08 | 1,04E+14 | 4,98E+08 | 5,00E+08 | 5,00E+08 |
| Failure rate lambda (1/h) | 1,00E-08 | 1,00E-14 | 2,01E-09 | 2,00E-09 | 2,00E-09 |
| FIT | 1,00E+01 | 9,60E-06 | 2,01E+00 | 2,00E+00 | 2,00E+00 |
| MDT (h) | 24 | 12 | 23.95 | 24 | 24 |
| MTTF (h) | 1,00E+08 | 1,00E+14 | 4,98E+08 | 5,00E+08 | 5,00E+08 |
| Mission time (h) | 8760 | 8760 | 8760 | 8760 | 8760 |

Table 4 shows the comparison between the expected values and actual values of the respective dependability properties. In this case study, the acceptable limit is set to the expected value due to simplicity. Normally, the comparison is done between the acceptable limit and the actual values. This comparison describes the fulfillment of the dependability goals. The expected value of failure rate is set to 10 fit, this value is used 5 times for comparison (5 corresponds to the number of measures). Compared with this value, the acceptance value of objective failure rates of different measures is obtained (e.g. 2 fit < 10 fit in Table 4). Afterwards, the subject evaluations will be performed. Such subjective evaluation offers additional but essential acceptance criterion. For example, if reliability or availability target values cannot be totally fulfilled, it is important to know what the drawbacks of non-fulfillment are and what would be the cost and time to achieve the target value. Based on the objective comparison and these subjective comparisons of the measures regarding the aforementioned acceptance criteria, the overall acceptance (e.g. 1: total acceptance in Table 4) will be subjectively determined.

**Table 4.** Objective and subjective evaluation of alternatives / measures (monitoring 1 fit)

| Solution | Measure monitoring 1 fit |
|---|---|
| Failure rate / Hazard rate | |
| Actual value (fit) | 2 |
| Expected value (fit) | 10 |
| Acceptable upper limit (fit) | 10 |
| Acceptable lower limit (fit) | |
| Evaluation of benefit of actual value | Better reliability of availability of the system |
| Evaluation of drawback of actual value | None |
| Cost of improvement towards expected value | None |
| Time-to-achievement of the improvement | |
| Overall acceptance | 1: total acceptance |

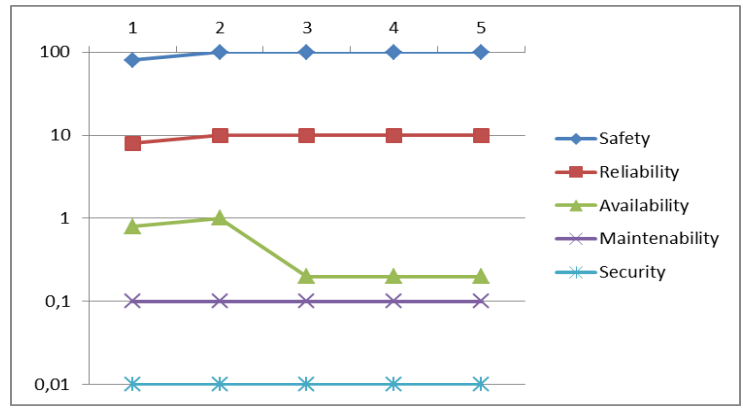

**Fig. 4.** Evaluation results according to the objective and subjective evaluation criteria

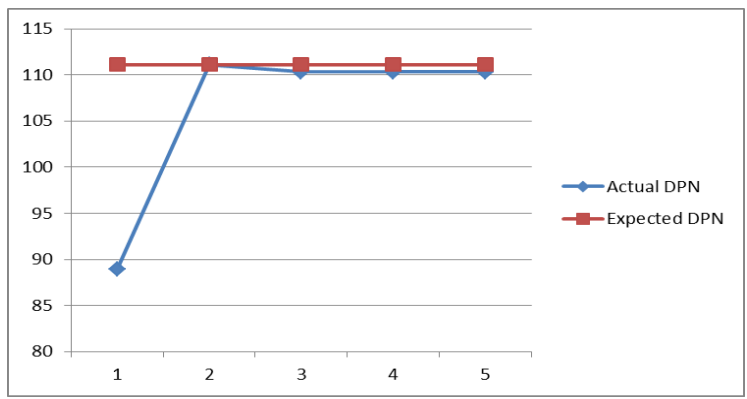

**Fig. 5.** Comparison of the actual DPN and expected DPN of alternatives / measures

**Table 5.** Dependability Priority Number of measures / alternatives

| Statistic | Without measure | Measure 1 (Redundancy) | Measure 2 (Monitoring: FR 1) |
|---|---|---|---|
| Safety | 80 | 100 | 100 |
| Reliability | 8 | 10 | 10 |
| Availability | 0,8 | 1 | 0,2 |
| Maintainability | 0,1 | 0,1 | 0,1 |
| Security | 0,01 | 0,01 | 0,01 |
| **DPN** | 88.91 | 111.11 | 110.31 |

Finally, the DPN is calculated based on this acceptance value and the respective weights of the properties according to $\sum_{k=i}^{n} X_{ij} * K_i$. For instance, the fifth measure of monitoring with failure rate of 1 fit fulfills the safety target value, but does not fulfill availability expected value (0,2 as shown in Table 5. and Fig. 4). Table 5. shows the results of $X_{ij} * K_i$. Therefore, the actual $DPN_{alternative1} \sum_{k=i}^{n} X_{i1} * K_1 = (100 + 10 + 0.2 + 0.1 + 0.01) = 110.31$ with $K_{safety} = 100$, $K_{reliability} = 10$, $K_{availability} = 1$ $et$. The expected $DPN_{alternative5} = \sum_{k=i}^{n} X_{i1} * K_1 = 100*1+10*1+1*1+0.1*1+0.01*1=111.11$. These two values are visualized in Fig. 5. as the fifth points of each of the lines. The expected values of the alternatives are plotted as brown points, while the actual values the blue points. Obviously, this measure does not fulfill all the target values. In contrary, the 2nd measure, redundancy measure fulfills all the dependability targets. It has the highest actual DPN. The actual DPN of this measure is on the same level as the expected DPN. The actual DPNs of other measures are lower than the expected DPNs (shown as blue points under brown points). Fig. 4. also shows the comparison of changes of the dependability properties. By this for instance a conflict between safety and availability is identified. By keeping the safety on the same high value, the availability goes down by monitoring with 10000 fit (3rd measure) dramatically. However, this conflict is not handled further, because the 2nd measure was chosen as solution. Otherwise a trade-off must be found and according to the changed DPNs the optimal alternative is selected. Basically the more important property wins. Through this case study, the strength of the DPN is illustrated. Quantitative dependability analysis (CFT) is thereby integrated into dependability trade-off analysis and vice verse. This combination improves the dependability of the system and reduces the cost of ignorable conflicts between the dependability goals.

## 5      Conclusion

This work illustrates how the concept of Dependability Priority Numbers (DPN) supports quantitively trade-off analyses. DPN helps to select of the optimal system design alternative or measure, in order to fulfill dependability goals. Dependencies and conflicts can be identified and resolved inherently by using this approach. DPN brings model-based dependability analysis and trade-off analysis together. An exemplary case study illustrates the concept and benefits of DPN. Our approach supports not only the quantitative trade-off analysis, but also extending model-based quantitative dependability analysis towards trade-off analysis.

DPN will be further developed both conceptually and according to tool support. More quantitative and detailed acceptance evaluation criteria, utilization of effective pre-selection algorithm in case of handling of large number of alternatives, calculation of object and subject acceptance values towards DPN in a more effective way will be investigated in the future.

### Acknowledgement

This work is supported by the Framework Programme for Research and Innovation Horizon 2020 under grant agreement n.732242 (DEIS).